\documentclass[twocolumn,noshowpacs,pre,preprintnumbers,superscriptaddress,amsmath,amssymb,floatfix]{revtex4}
\usepackage[caption=false]{subfig}
\usepackage{graphicx}
\usepackage{color}
\usepackage{placeins}
\usepackage{tikz}
\usepackage{mathtools}
\usepackage{amsthm}
\usepackage{romannum}
\usepackage{color}
\usepackage{xkeyval,xcolor}

\newcommand{\kk}{\langle k \rangle}
\newcommand{\er}{Erd\H{o}s-R\'{e}nyi}

\newcommand{\subfigimg}[3][,]{%
	\setbox1=\hbox{\includegraphics[#1]{#3}}
	\leavevmode\rlap{\usebox1}
	\rlap{\hspace*{30pt}\raisebox{\dimexpr\ht1-2\baselineskip}{#2}}
	\phantom{\usebox1}
}

\begin{document}
	
	\title{Spreading of localized attacks in spatial multiplex networks}
	\author{Dana Vaknin}
	\affiliation{Bar-Ilan University, Ramat Gan, Israel}
	\author{Michael M. Danziger} 
	\affiliation{Bar-Ilan University, Ramat Gan, Israel}
	\author{Shlomo Havlin}
	\affiliation{Bar-Ilan University, Ramat Gan, Israel}
	\date{\today}
	
	\begin{abstract}
		
		Many real-world multilayer systems such as critical infrastructure are interdependent and embedded in space with links of a characteristic length.
		They are also vulnerable to localized attacks or failures, such as terrorist attacks or natural catastrophes, which affect all nodes within a given radius.
		Here we study the effects of localized attacks on spatial multiplex networks of two layers.
		We find a metastable region where a localized attack larger than a critical size induces a nucleation transition as a cascade of failures spreads throughout the system, leading to its collapse.
		We develop a theory to predict the critical attack size and find that it exhibits novel scaling behavior.
		We further find that localized attacks in these multiplex systems can induce a previously unobserved combination of random and spatial cascades.
		Our results demonstrate important vulnerabilities in real-world interdependent networks and show new theoretical features of spatial networks.
	\end{abstract}
	\maketitle
	
	\section{INTRODUCTION}
	\begin{figure}
		\centering
		\includegraphics[width=\linewidth]{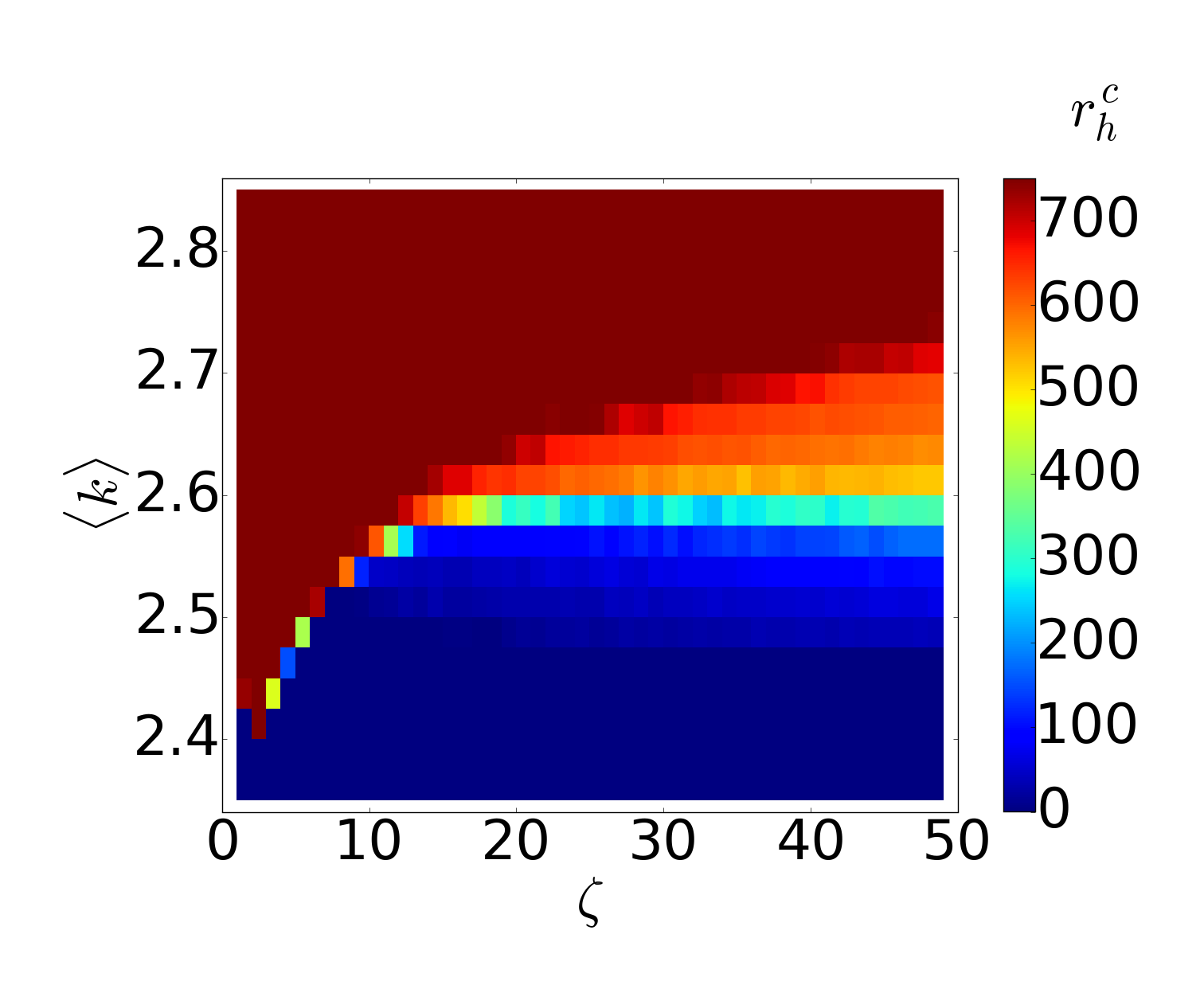}
		\caption{\textbf{Phase diagram of the critical attack size $r_h^c$.}
			Dependence of the critical attack size $r_h^c$ on the average degree $\kk$ and the characteristic length $\zeta$. The color bar in the right represent the size of $r_h^c$.
			In this figure $L = 1500$, averaged over 5 runs for each data point.}	
		\label{fig:rck}
	\end{figure}
	
	The important subject of vulnerability of complex systems has garnered much interest for many years.
	Most infrastructure is embedded in space, for example the power grid and sewer networks, and thus, several models have been proposed for understanding the vulnerability of spatially embedded networks
	\cite{doar1993bad,wei1993comparison,zegura1997quantitative, watts-nature1998,penrose2003random,kleinberg2000small,kosmidis-epl2008,li-naturephysics2011,mcandrew-pre2015}.
	In addition, in recent years, world-wide human, technological, social and economic systems have become more and more integrated and interdependent \cite{peerenboom-proceedings2001,rinaldi-ieee2001,rosato-criticalinf2008,bookstaber-ofr2016,majdandzic-naturecomm2016,li-preprint2014}.
	Therefore, it is necessary to realistically model these systems as interdependent in order to understand their structure, function and vulnerabilities \cite{buldyrev-nature2010,gao-naturephysics2012,gao-prl2011,baxter-prl2012,kivela-jcomnets2014,dedomenico-prx2013,bianconi-pre2013,boccaletti-physicsreports2014,gao-nationalscirev2014,cai-naturephysics2015,hu-prx2014,danziger-collection2016}.
	Studies on spatially embedded interdependent networks found that in many cases they are significantly more vulnerable than non-embedded systems \cite{barthelemy-physicsreports2011,bashan-naturephysics2013,wei-prl2012,danziger-jcomnets2014,boccaletti-physicsreports2014,shekhtman-pre2014,asztalos-plosone2014,danziger-newjphysics2015,lee-epjb2015}.
	
	Though most research on resilience of complex systems considers random failures, in many cases, nodes fail in localized areas, due to natural catastrophes, terrorist attack or other failures.
	Recent studies show that localized attacks on some systems are significantly more damaging
	\cite{neumayer-milcom2008,agarwal-infocom2011,berezin-scireports2015,shao-njp2015,yuan-pre2015,yuan-pre2016,yan-springerbook2016,hu-scireports2016,wu-eng2016,ouyang-eng2016}.
	
	\begin{figure*}
		\centering
		\begin{tabular}{@{}p{0.33\linewidth}@{\quad}p{0.33\linewidth}@{\quad}p{0.33\linewidth}}
			\subfigimg[width=\linewidth]{(a)}{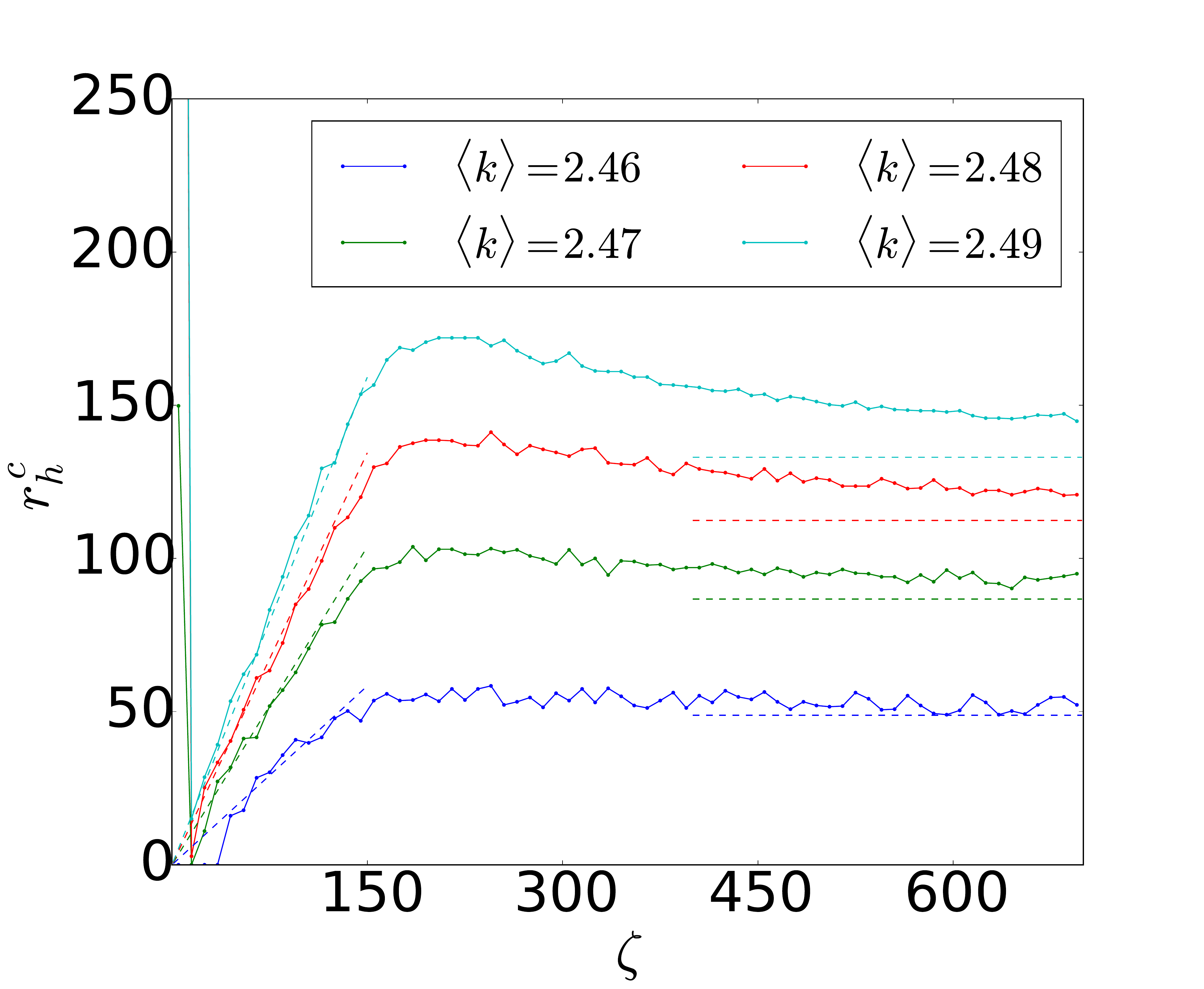} &
			\subfigimg[width=\linewidth]{(b)}{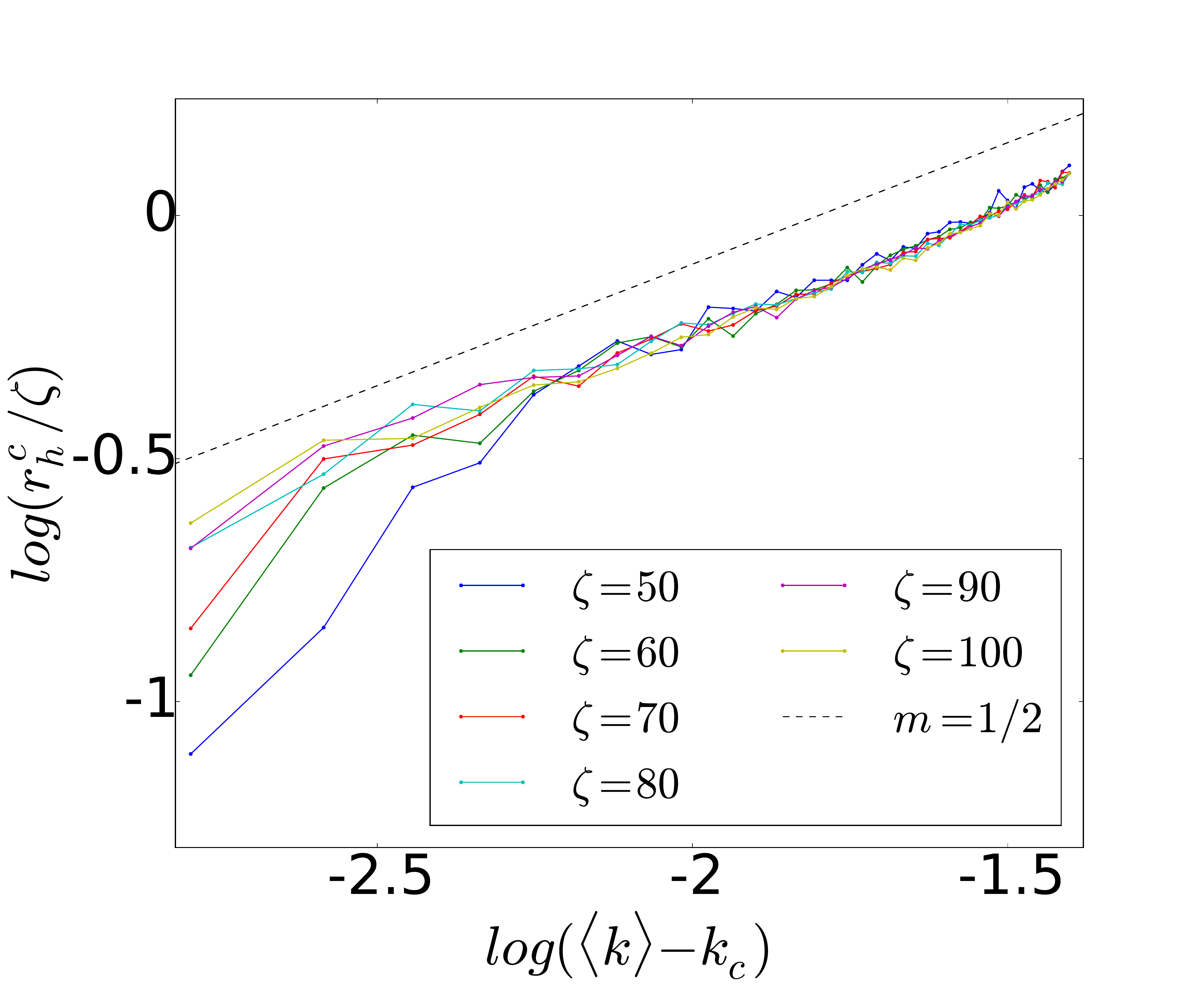} &
			\subfigimg[width=\linewidth]{(c)}{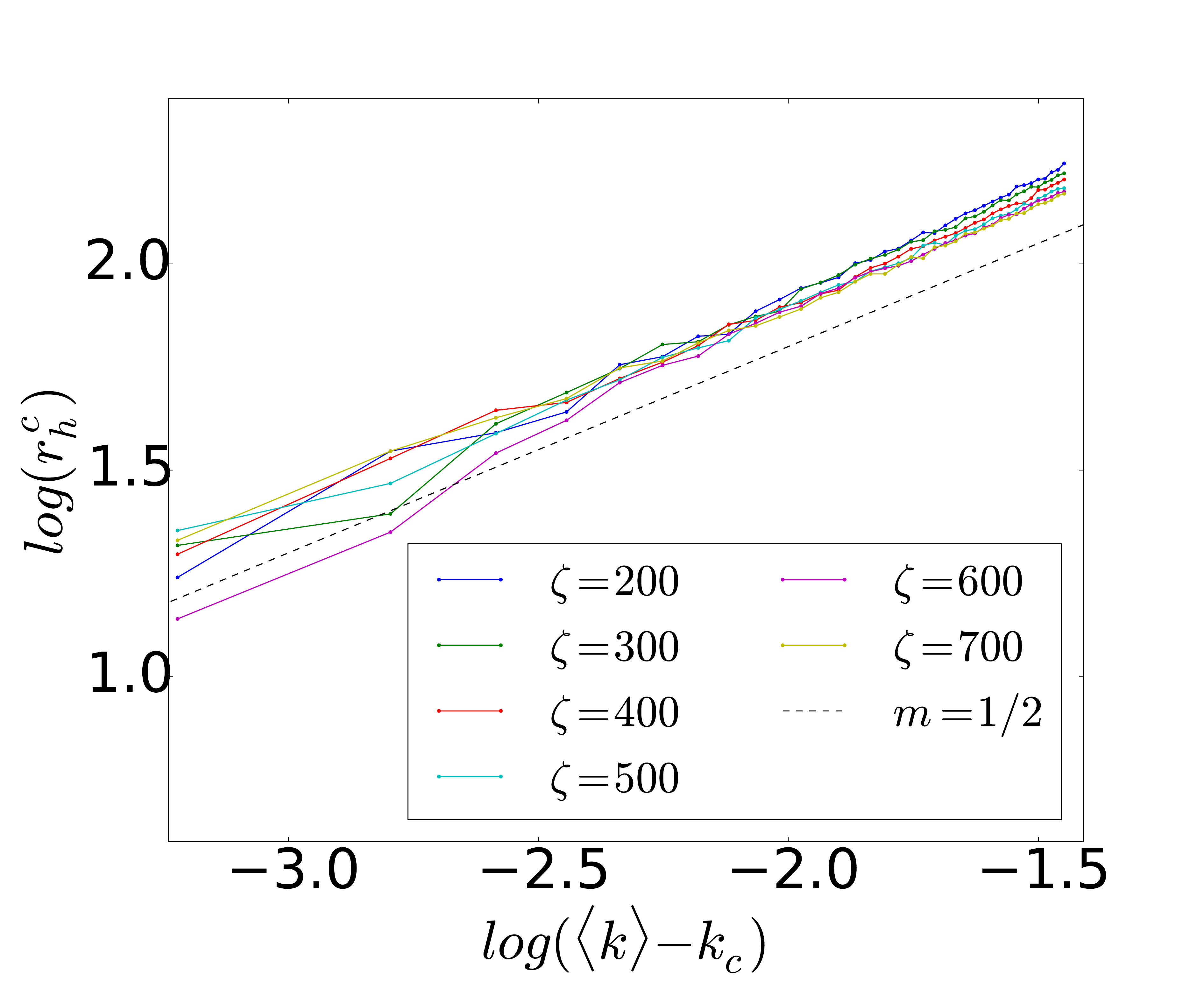} 
		\end{tabular}
		\caption{\textbf{The critical attack size $r_h^c$ --- simulations and theory.} 
			(a) $r_h^c$ as a function of $\zeta$ for four $\kk$ values. The dotted lines represent the theory for small and large $\zeta$ as obtained from Eqs. (\ref{eq:small_zeta_exact}) and (\ref{eq:large_zeta_exact}) respectively.
			(b) $\log(r_h^{c}/\zeta)$ and (c) $\log(r_h^{c})$ as a function of $\log(\kk - k_c)$ for small and large $\zeta$ values, with the $\frac{1}{2}$ exponent (dotted line), predicted by the theory (see Eqs. (\ref{eq:small_zeta_approximation}) and (\ref{eq:large_zeta_approximation})).
			For this figure $L = 2000$ with averages over at least 5 runs for each data point.} 
		\label{fig:big_small_zeta}
	\end{figure*}
	
	Here, we study localized attacks on a realistic spatial multiplex model that has been proposed recently \cite{danziger-epl2016,grossvaknin-JPS}. The system is a model of multiplex  with exponential link-length distribution of connectivity links in each of the two layers:
	\begin{equation}
	P(r) \sim \exp(-\frac{r}{\zeta}) 
	\label{eq:exponential_distribution}.
	\end{equation} 
	Here $\zeta$ is a parameter determining the characteristic link length and thereby the strength of the embedding --- a smaller $\zeta$ reflects a stronger embedding.
	We further assume that the nodes require connectivity in each layer in order to function, a requirement which is equivalent to having dependency links of length zero with longer connectivity links.
	This is in contrast to the research based on the model of Li et al. \cite{wei-prl2012} and Berezin et al. \cite{berezin-scireports2015} which considered the case where dependency links are longer then connectivity links.
	We suggest that the assumption of dependency links which are shorter than connectivity links is more natural, because, for example, it is more likely for a communication's station to receive power from its nearest power station than a distant one, though the communications and power networks are known to have potentially long links \cite{danziger-epl2016,saleh-arxiv2015}.
	
	As we show here, the combination of spatially constrained connectivity links and multiplex dependency---both ubiquitous features of real complex systems---makes these systems vulnerable to potentially catastrophic localized attacks. 
	Such attacks are important and realistic because they can represent a local damage on two spatial networks that depend on one another to function in a very natural way: the nodes are either the same, or every node in one network layer depend on a close node in the other.
	
	We find that for a broad range of parameters our system is metastable, meaning that a localized attack larger than a critical size --- that is independent of the system size --- induces a cascade of failures which propagates through the whole system leading to its collapse.
	We develop a theory which can predict this critical size of the initial local damage, and can explain the unique cascading process that makes the critical size independent of the system size.
	We find that when the localized attack is of the critical size --- the cascade is at first random within a disc of radius of order $\zeta$, and then it propagate  spatially until it reaches the boundaries of the system.
	Using this theory we also find a new scaling exponent describing the critical nucleation (of damage) size.

	\section{MODEL}
	
	We model the multiplex composed of two layers in which the nodes are placed at lattice sites of a square lattice where the link lengths $r$ are distributed with probability of Eq. (\ref{eq:exponential_distribution}) and average degree $\kk$. 
	Here, we focus on the case in which both layers have the same characteristic length $\zeta$ and same $\kk$. 
	In practice, we assign each node an $(x,y)$ coordinate with integers $x,y \in [0,1,...L)$, and construct the links in each layer as follows:
	(a) We select randomly a source node $(x_s,y_s)$ and draw an angle $\alpha$ selected uniformly at random.
	(b) We draw a length $r$ selected from the distribution $P(r)$, Eq. (\ref{eq:exponential_distribution}).
	(c) We select the target node $(x_t,y_t)$, which is closest to satisfying, $(x_t,y_t) = (x_s,y_s) + (r\cdot \cos\alpha,r\cdot \sin\alpha)$.
	This process is executed independently in each layer and is continued until we have a total of $\frac{N\kk}{2}$ links.
	The topological model is similar to the Waxman model \cite{waxman-ieee1988} and recent work by Bianconi and Halu et al.~\cite{bianconi-pre2013,halu-pre2014} with a key difference being that our model converges to a lattice as $\zeta\to 0$.
	
	For a node to remain functional it must be connected to the giant component in both layers. 
	This reflects the assumption that in order for the node to continue to function it requires the two types of connectivity.
	Next, we perform a localized attack as follows:
	(a) We remove all nodes within a distance $r_h$ from a random location in the system.
	(b) From the set of the remaining nodes, we remove all the nodes that are not in the giant component of the first layer.
	(c) We repeat step (b) in the second layer.
	(b) and (c) are repeated until there are no nodes to remove, and we are left with the mutual giant component (MGC) \cite{buldyrev-nature2010,gao-naturephysics2012,bianconi-pre2015,cellai-pre2016}.
	
	At the end of this cascade, the system is categorized as functional or non-functional depending on whether the MGC is of the order of the system size $L^2$ or not.
	
	\section{RESULTS}
	
	\begin{figure*}
		\centering
		\subfloat[]{\includegraphics[width=0.34\linewidth]{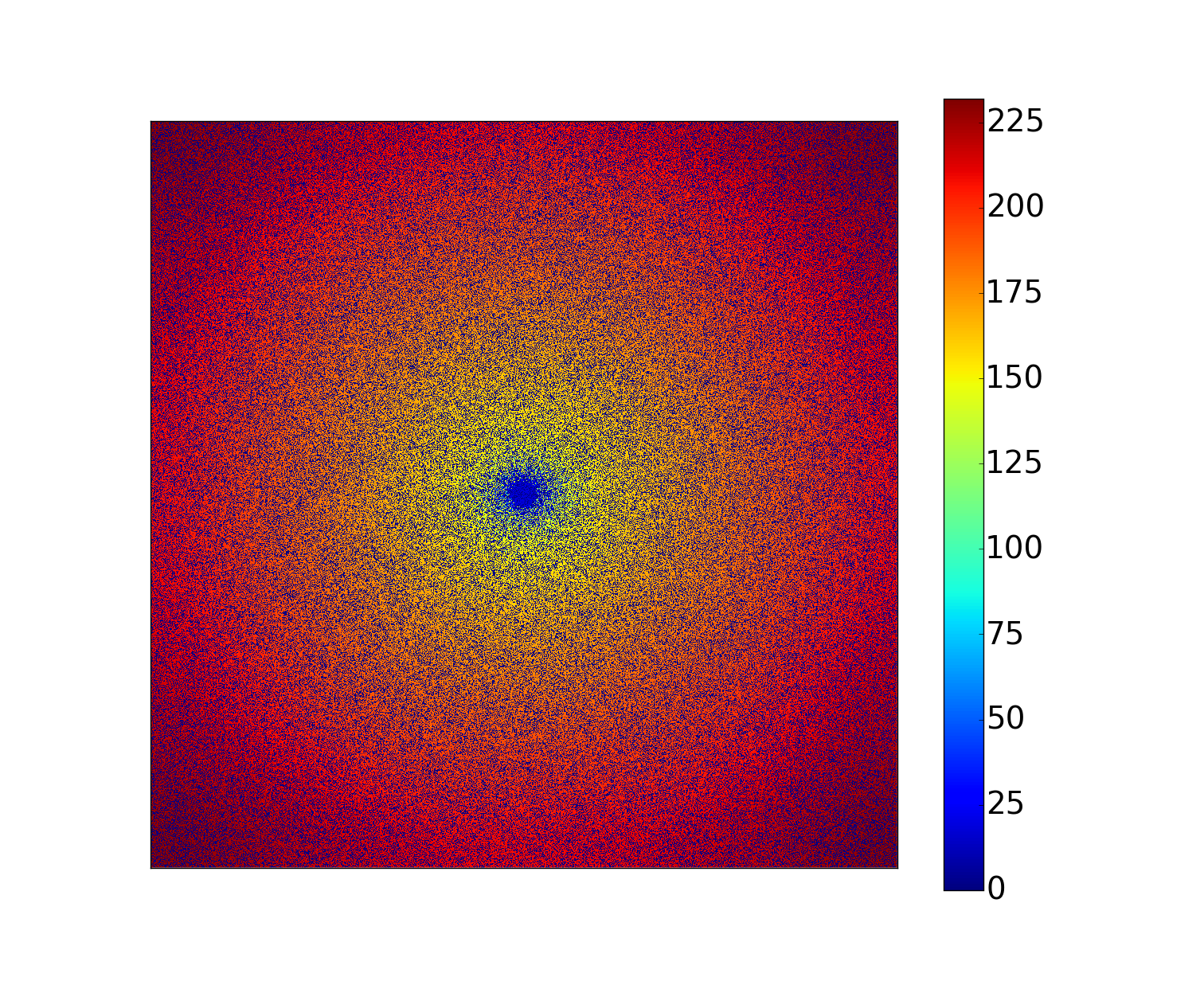}}\hfil
		\subfloat[]{\includegraphics[width=0.33\linewidth]{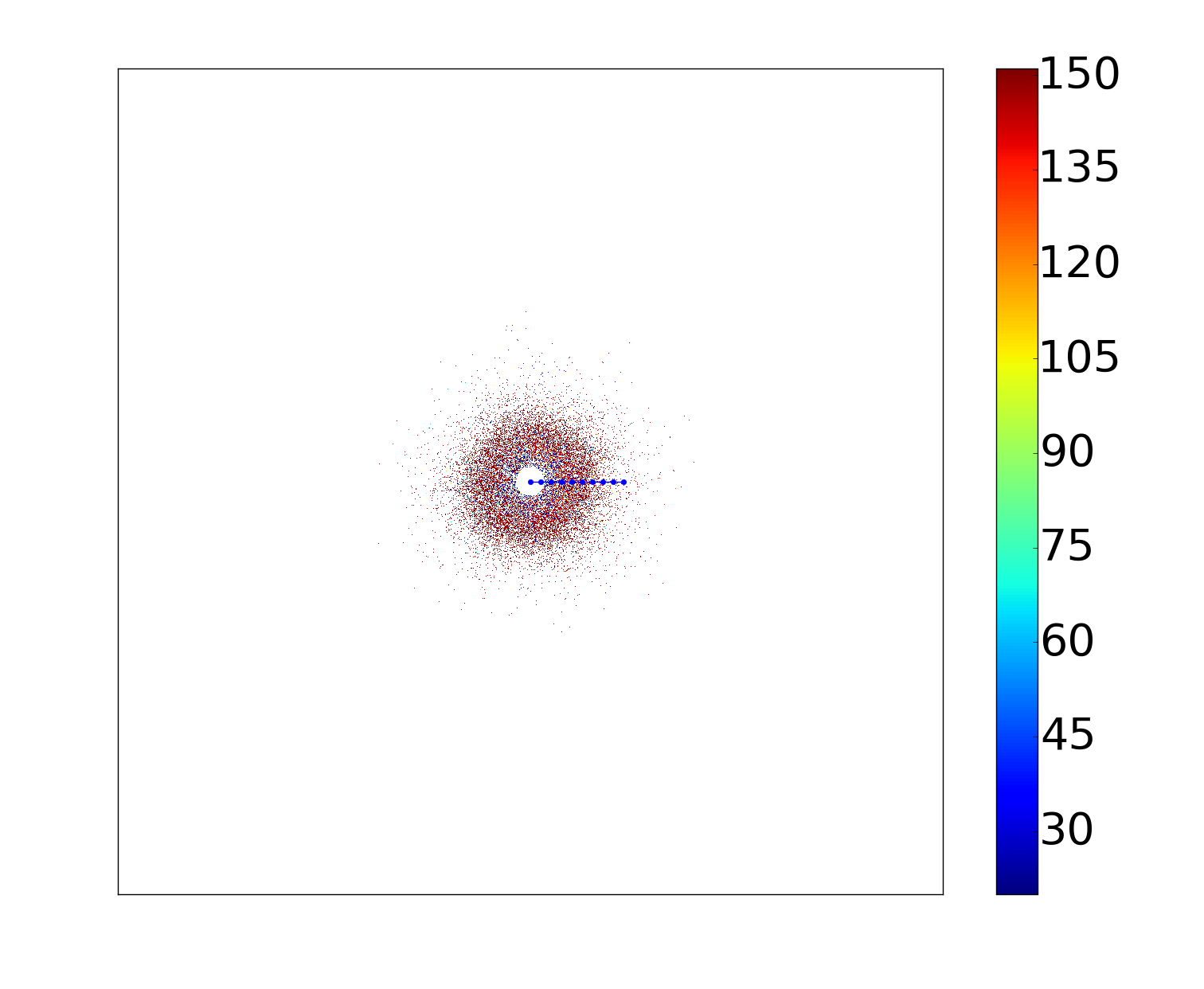}}\hfil
		\subfloat[]{\includegraphics[width=0.33\linewidth]{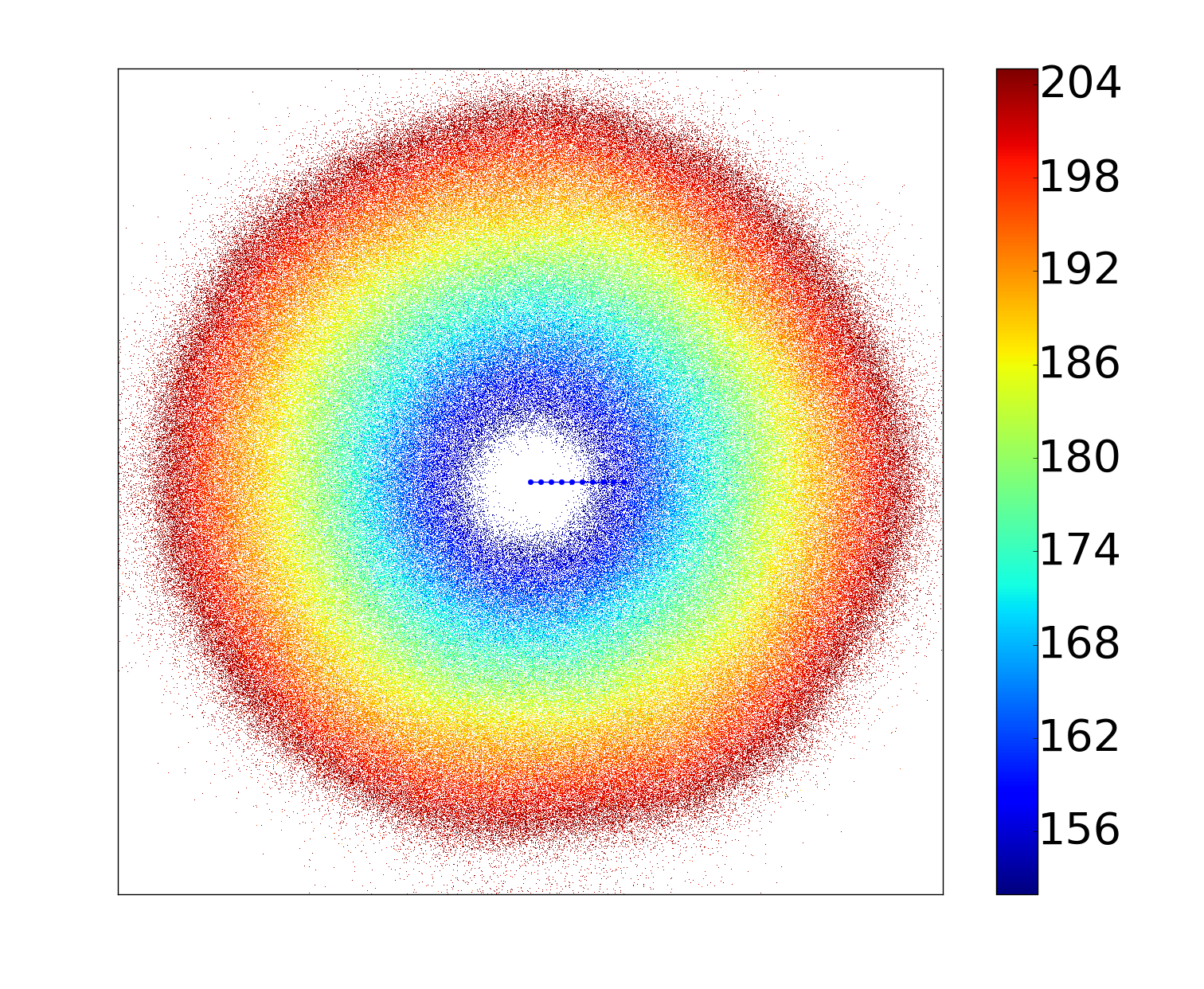}}
		\caption{\textbf{Dynamic evolution of cascading failures near the critical point.} 
			Propagation of local damage with radius slightly above the critical size $r_h^c$. The colors represent the number of iterations (NOI) until the nodes fail. In (b) and (c) we show a ruler in $\zeta$ units (9$\zeta$) for demonstrating the orders of magnitude.
			For this figure $L = 4000$, $\zeta = 50$, $\kk = 2.5$ and $r_h = 69$.}
		\label{fig:age}	
	\end{figure*}
	
	We analyzed the damage spreading of the localized attack on the multiplex with different $\kk$, $\zeta$ and $r_h$. Our simulations suggest the existence of $r_h^c$, a minimum radius of damage needed to cause the system to collapse. 
	Below $r_h^c$ the damage remains localized while for a radius above $r_h^c$ the damage propagates indefinitely and destroys the whole multiplex.
	When we calculate the critical attack size $r_h^c$ for different $\kk$ and $\zeta$, we discover three regions, as shown in the phase diagram in Fig.\ref{fig:rck}. 
	The regions are:
	(a) Stable (in red) --- in this region the system remains functional after a localized attack of any finite size.
	(b) Unstable (in blue) --- in this region the system is non-functional even if no nodes are removed.
	(c) Metastable (between the above-mentioned regions) --- in this region only attacks with radius larger than $r_h^c$ propagate, through cascading failures, the entire system and makes it non-functional.
	
	To understand these phenomena we consider the network as being composed of regions of size of order $\zeta$ that are tiled on a 2D lattice, each of which can be approximated as a random network. 
	The localized attack of size $r_h$ can then be approximated as a random attack of size $\pi{r_h}^2$ in an interdependent random network with $\sim \zeta^2$ nodes.
	In this case, the percolation threshold for random removals is known to be $p_c = \frac{k_c}{\kk}$, where $k_c \approx 2.4554$ \cite{buldyrev-nature2010}.
	It has also been shown that localized attacks (formed by shells surrounding a root node) in \er~multiplex networks have the same percolation threshold as random attacks \cite{shao-njp2015,yuan-pre2015}.
	
	Based on this, we can predict the critical attack size $r_h^c$ close to $k_c$ as follows:
	\begin{equation}
	\frac{\pi(r_{h}^{c})^{2}}{\pi(a\zeta)^2} \cong 1 - \frac{k_c}{\kk},
	\label{eq:small_zeta_exact}
	\end{equation}
	from which,
	\begin{equation}
	\frac{r_{h}^{c}}{\zeta} \cong a \cdot \sqrt{1 - \frac{k_c}{\kk}} \cong \frac{a}{k_c} \cdot \sqrt{\kk - k_c}
	\label{eq:small_zeta_approximation},
	\end{equation}
	where $a$ is the constant of proportionality for the effective random network (radius) size, which we determine numerically.
	This $r_{h}^{c}$ is the minimal expected size of the hole that destroys the entire random network regime ($a\zeta$). 
	However, since there are links between the tiled \er~sub-networks, the collapse propagates toward the surrounding sub-networks and we see a typical spreading cascade in an embedded network.
	
	For the limit of $\zeta$ of the order of $L$, the multiplex can be well approximated as two interdependent \er~networks, and therefore we can calculate $r_h^c$ as follows:
	\begin{equation}
	\frac{\pi (r_{h}^{c})^{2}}{L^2} \cong 1 - \frac{k_c}{\kk},
	\label{eq:large_zeta_exact}
	\end{equation}
	from which,
	\begin{equation}
	r_{h}^{c} \cong \frac{L}{\sqrt{\pi}} \cdot \sqrt{1 - \frac{k_c}{\kk}} \cong \frac{L}{\sqrt{\pi k_c}} \cdot \sqrt{\kk - k_c}\label{eq:large_zeta_approximation}.
	\end{equation}
	
	We show that Eq.(\ref{eq:small_zeta_exact}) and Eq.(\ref{eq:large_zeta_exact}) predict the simulation results in Fig. \ref{fig:big_small_zeta}(a) with $a \approx 9$.
	Because of the long links and since the sub-networks are not isolated ---
	$a$ is relatively big.
	In the supplementary we can see similar phenomenon on a system with average degree $\kk$ with links that connected slightly different. In this alternative model we choose a node randomly and link it to another node with link-length distribution of step function up to $\zeta$, for each of the two layers.
	In this case, there are no long links (but the small \er~networks are still not isolated) so $a$ is found to be smaller than in our model (approximately 3.2).
	
	Eq.(\ref{eq:small_zeta_approximation}) and Eq.(\ref{eq:large_zeta_approximation}) also predict a novel critical exponent for the scaling of $r_h^c$ with $\kk - k_c$, which equals $\frac{1}{2}$. Indeed the simulations shown in  Fig. \ref{fig:big_small_zeta}(b) and (c) support this exponent. 
	Generally, it is difficult to find evidence for universality in the absence of a second-order transition. This new scaling, related to nucleation type processes, may provide an alternative approach which can be useful to understand universality properties in critical phenomena associated with a first-order transition where nucleation processes are involved \cite{mcgraw-prl1996,talanquer-jcp1997}.
	
	\begin{figure}
		\includegraphics[width=0.87\columnwidth]{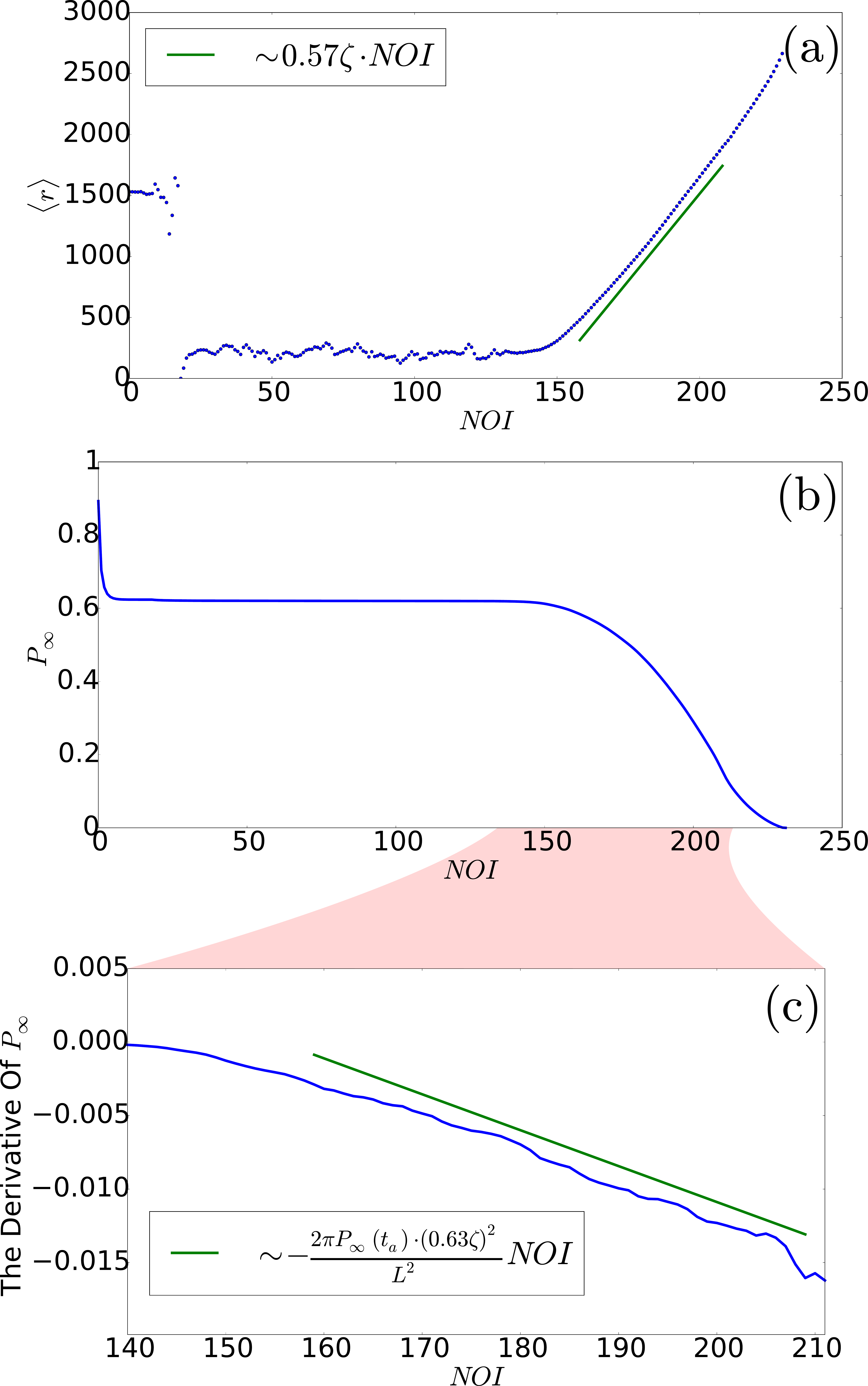}
		\caption{\textbf{Analysis of the cascading failures near the critical point.} 
			(a) The average distance from the center, $\langle r \rangle$, of the nodes that fail at every iteration, with a linear fit for the spatial spreading phase. 
			(b) The size of the MGC, $P_\infty$, as a function of NOI.
			(c) The derivative of $P_\infty$ with comparison in the spatial spreading process to Eq. (\ref{eq:diff_pinf}).
			For this figure $L = 4000$, $\zeta = 50$, $\kk = 2.5$ and $r_h = 69$, the same runs as Fig. \ref{fig:age}.}
		\label{fig:age3}
	\end{figure}
	
	\begin{figure}
		\includegraphics[width=\columnwidth]{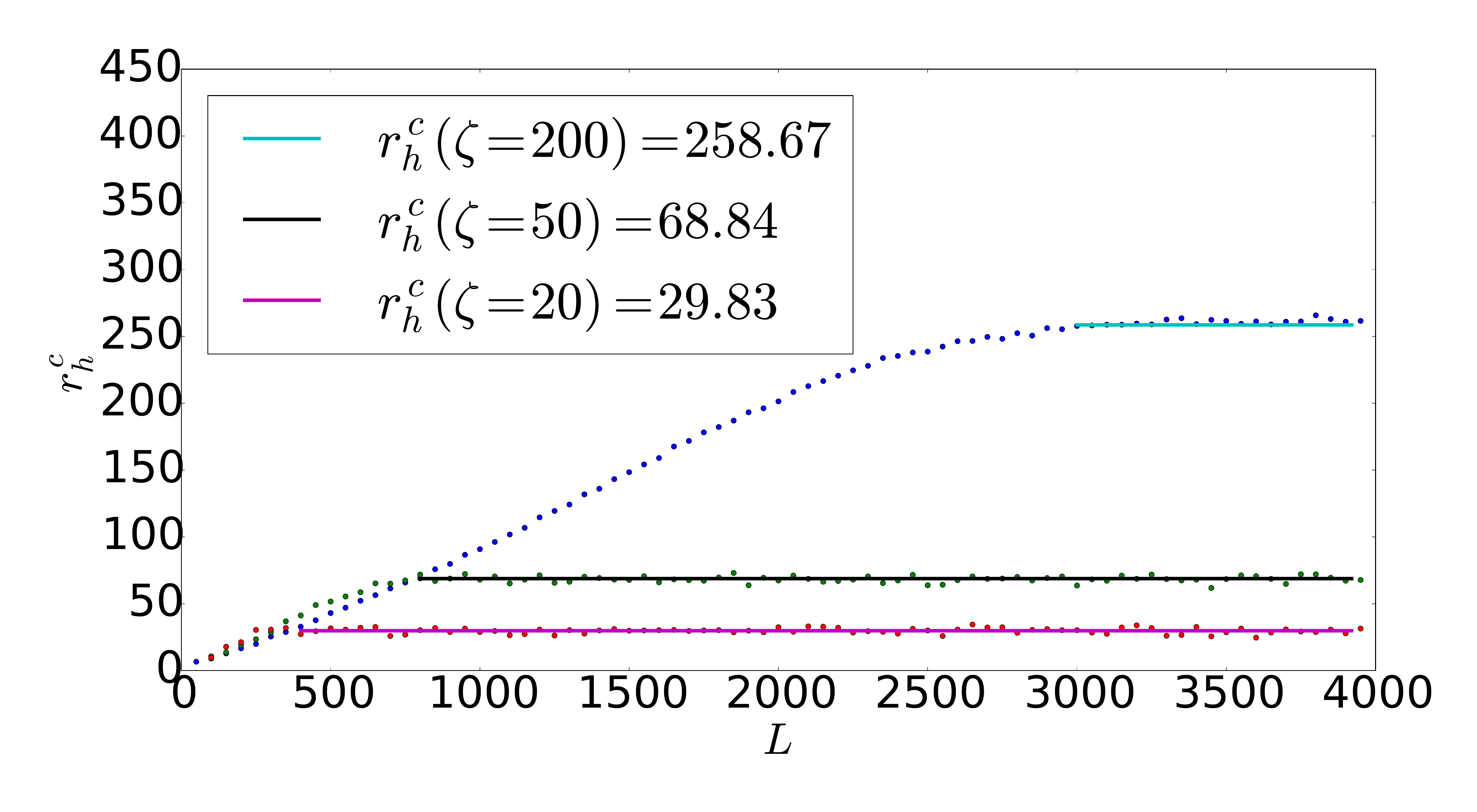}
		\caption{\textbf{Dependence of the critical attack size $r_h^c$ on the system size $L$.} 
			We see that above a certain value of $L$ the critical attack size $r_h^c$ is constant.
			For this figure $\kk = 2.5$, with 5 runs for each data point.}	
		\label{fig:rh_L}
	\end{figure}
	
	We also find a new dynamical process of cascading when the localized attack is near the critical size, that is consistent with our theory.
	To understand this process for a given $\kk$ and $\zeta$, we first find the MGC, and then we remove a hole with radius $r_h^c$ needed to initiate the cascading.
	Fig. \ref{fig:age}(a) shows the whole spatial-temporal process of cascading and Figs. \ref{fig:age} (b) and (c) demonstrate the different types of NOI in the two regimes as described below.
	The graph in Fig. \ref{fig:age3}(a), of $\langle r \rangle$, the average distance from the center of the nodes that failed in every iteration, reveals explicitly the three main stages of the whole process shown in Fig. \ref{fig:age}(a):
	(\romannum{1}) Before the localized attack, there are a few steps where the cascade describes the removal of nodes that are not in the MGC, so $\langle r \rangle$ is close to the average distance from the center to all nodes ($\sim 1500$).
	(\romannum{2}) Random branching process \cite{dong-pre2014,lee-pre2016} in limited annulus around the hole (demonstrated in Fig. \ref{fig:age} (b)) so $\langle r \rangle$ is fixed for many iterations at distance $\approx \frac{a}{2} \cdot \zeta$.
	(\romannum{3}) Spatial spreading process that propagates the whole system (demonstrated in Fig. \ref{fig:age} (c)), so $\langle r \rangle$ increases linearly as a function of number of iterations (NOI). 
	Indeed we can see the effect of the three above processes in Fig. \ref{fig:age3}(b) --- the size of the MGC, $P_\infty$, at first decreases sharply, then, after the attack in $t_a$, it decreases very slowly in a plateau, and then parabolically  as a function of NOI.
	Additionally, the processes are also described in the supplementary on the discussion about the branching factor.
	
	In the spreading process we can see the cascading dynamics both in the simulations for $\langle r \rangle(t)$ and in the simulations for $\frac{\partial P_\infty(t)}{\partial t}$ in Fig. \ref{fig:age3}(c). The connection between them is expressed in the equations below so that $t$ expresses the NOI and $v$, that sets the speed of the cascading, is $ \sim 0.6 \zeta$,
	
	\begin{gather}
	\langle r \rangle(t) \cong  vt + r_h^c \\
	\frac{\partial P_\infty(t)}{\partial t} \cong -\frac{P_\infty(t_a)}{L^2} \cdot \frac{\partial {[\pi(vt + r_h^c)^{2}-\pi {r_h^c}^{2}]} }{\partial t} \\
	\frac{\partial P_\infty(t)}{\partial t} \cong - \frac{2 \pi P_\infty(t_a) v^2}{L^2} \cdot t - \frac{2\pi P_\infty(t_a) v r_h^c}{L^2}.
	\label{eq:diff_pinf}
	\end{gather}
	
	Understanding the dynamical process of cascading can explain why in the metastable region, when the size of the network crosses the size of our approximated random network (around point $L\approx2a\zeta$ in Fig. \ref{fig:rh_L}), there is no correlation between the critical attack size $r_h^c$ and the system size.
	This is because once the network is large enough for a damage spreading process to take place, the hole will spread until the damage reaches the edges of the system, regardless of its size.
	
	\section{Discussion}
	We have presented a study of interdependent spatial networks with a novel and realistic combination of spatially localized damage and connectivity links which are longer than the dependency links. 
	This combination is ubiquitous in nature, and yet has not been studied  methodically, to our knowledge.
	We find that a nucleation phenomenon can be triggered by local damage, with failures spreading through the entire system.
	The cascade itself has multi-universal behavior---random on a small scale but spatial on a large scale---complementing the static bi-universality in the single layer case \cite{grossvaknin-JPS}.
	We further find that the critical nucleation size has novel scaling features.
	Future research will determine whether this indicates a general, universal feature of nucleation transitions.
	
	\section*{Acknowledgement}
	The authors acknowledge the Israel Science Foundation, Israel Ministry of Science and Technology (MOST) with the Italy Ministry of Foreign Affairs, MOST with the Japan Science and Technology Agency, ONR and DTRA for financial support. MMD thanks the Azrieli Foundation for the award of an Azrieli Fellowship grant.
	
	\bibliographystyle{unsrt}
	\bibliography{mybib}
	
	\newpage
	\widetext
	\section*{Supplementary}
	\appendix
	\section{Branching Process}
	
\begin{figure}
		\includegraphics[width=0.5\linewidth]{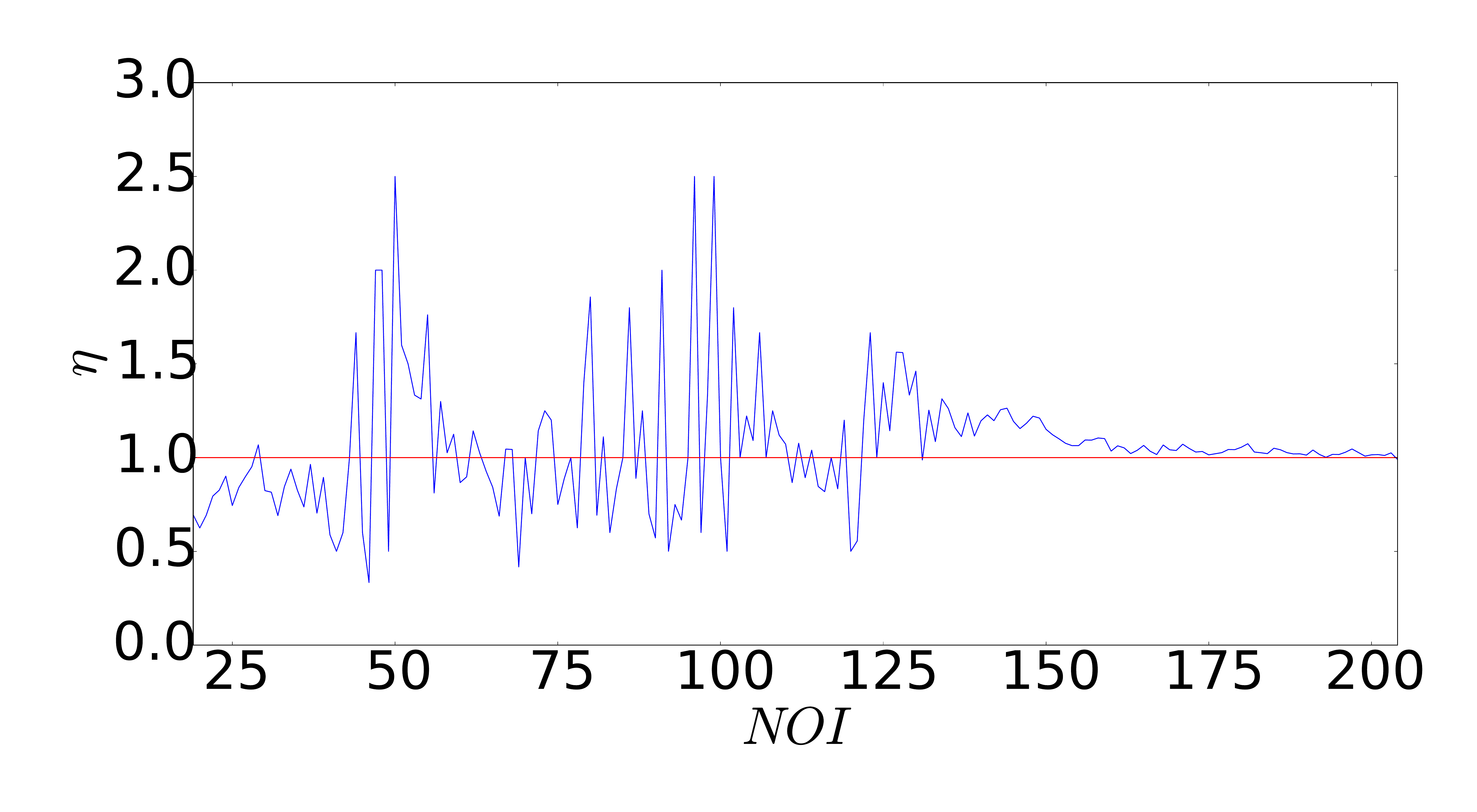}
		\caption{\textbf{The branching factor $\eta$}
			in the random branching process ($NOI<125$) and in the spatial spreading process ($NOI>125$), with $\eta = 1$ marked in red.
			In this figure $L = 4000$, $\zeta = 50$, $\kk = 2.5$ and $r_h = 69$ (same run as in Figs. 3-4 in the main text).}
		\label{fig: etta2}
		
	\end{figure}

	The branching factor $\eta$ is defined as the ratio between the number of nodes that collapse at a given iteration to the nodes that collapsed in the previous iteration. 
	In Fig. \ref{fig: etta2} we see that $\eta$ exhibits different behavior in the two main processes that occur after the attack.
	In the initial random-like stage, when the nodes fall slowly in the annulus around the hole, $\eta$ is noisy with an average value of one, meaning the number of the nodes that fail in every iteration is the same on average, though with considerable deviations, as observed in random networks~\cite{dong-pre2014}. 
	Then, when the damage propagates in a spreading cascade, $\eta$ is consistently above one ($NOI > 125$), meaning that the number of nodes that fail in every iteration is growing. There is a decreasing trend in $\eta$, reflecting the flattening of the interface (Fig. 3(c) in main text) as its radius increases. In the limit of an infinite radius of curvature (i.e., a propagating strip), the number of nodes failing in each step will be nearly identical to the previous step and $\eta \approx 1$.
	
	These results strengthen our claim that, following a critical localized attack, there is a random cascade followed by a spatial cascade.
	At first we have the branching process phenomena that characterize the critical cascade of an \er~multiplex, after which we see that the damage rate reflects collapse propagation like a typical spreading process in a spatial system.
	
	\section{Uniform link-length distribution with strict length cut-off at $\zeta$}
	
	\FloatBarrier
	
	\begin{figure}
		\includegraphics[width=0.39\linewidth]{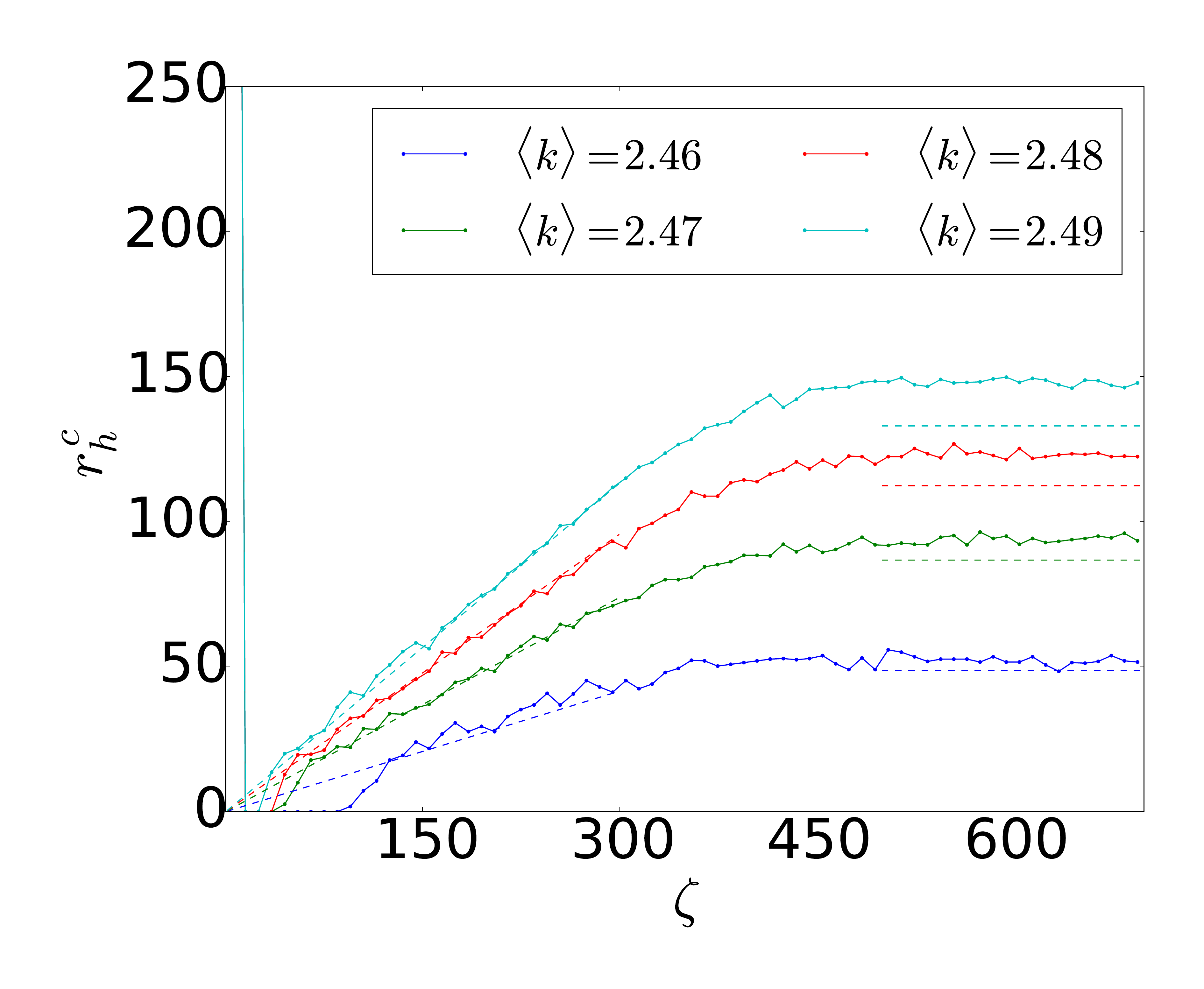}
		\caption{\textbf{The critical attack size $r_h^c$} 
			as a function of $\zeta$ for four $\kk$ values, with the fit from the theory for large $\zeta$ and small $\zeta$ with $a = 3.2$.
			For these simulations, $L = 2000$ with averages over 5 realizations for each data point.}
		\label{fig: rh_zeta2}	
	\end{figure}
	
	The model in the main text had exponential link-lengths distribution which was found to be well approximated as uniformly random up to the characteristic length ($\zeta$) multiplied by a constant $a\approx 9$.
	We thus hypothesize that similar random-like topologies tiled on a lattice will demonstrate similar behavior, with possible corrections to the constant $a$.
	Here we consider a multiplex composed of two layers in which the nodes are placed at lattice sites of a square lattice where the average degree is $\kk$ and the link length distribution is uniform up to $\zeta$ (step function, i.e., no links above length $\zeta$).
	
	In Fig. \ref{fig: rh_zeta2}, we see the results for this system which are explained with the same theory as the model in the main text, but with a modified coefficient $a$.
	In this case, there are no long links as in the model in the main text, so $a$ --- the constant of proportionality for the effective network size --- is significantly smaller, and from numerical calculations we find $a \approx 3.2$.
	The length $a\cdot \zeta$ can be regarded the effective radius of the random network.
	
	\begin{figure}
		\includegraphics[width=0.39\linewidth]{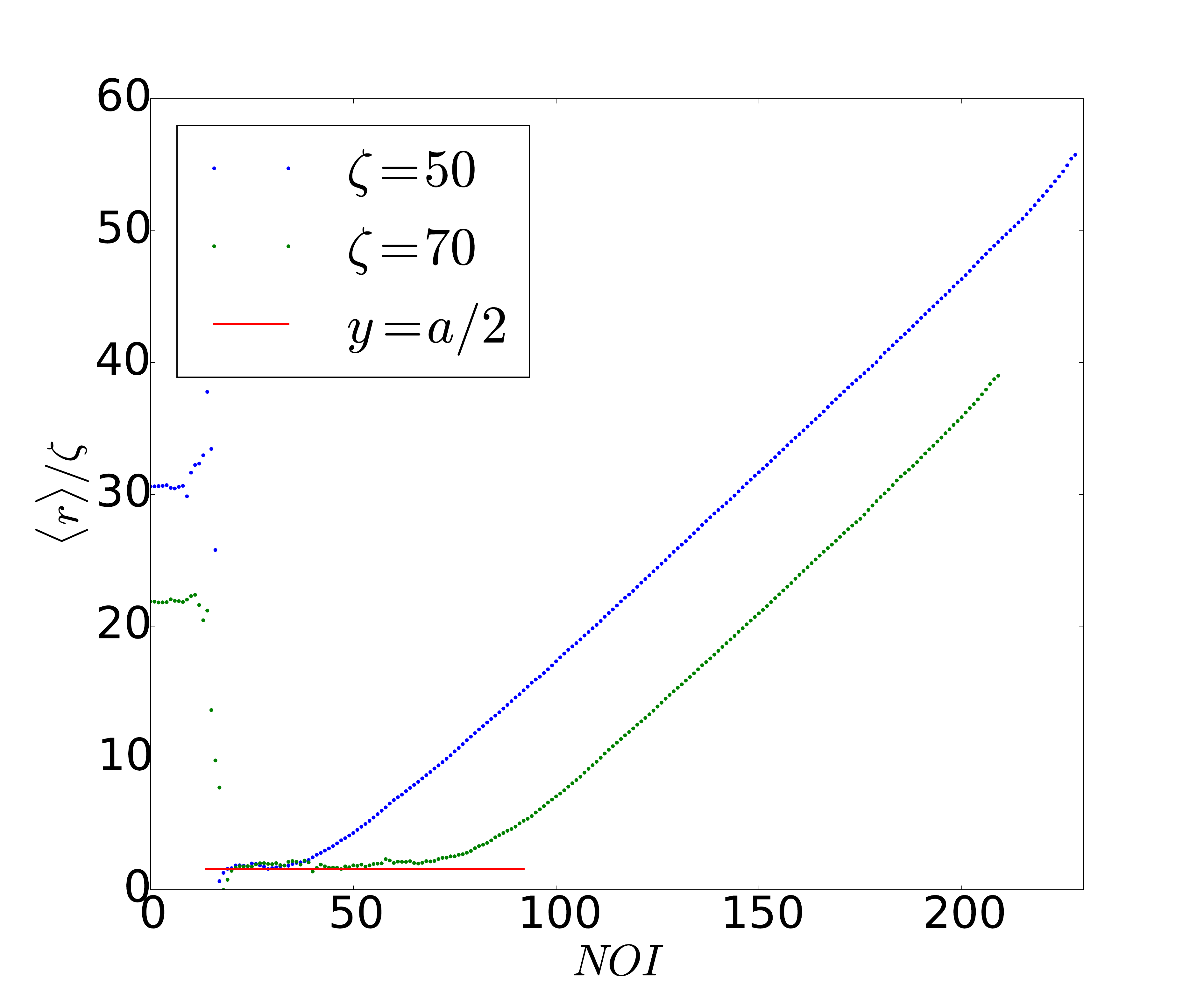}
		\caption{\textbf{The average distance from the center $\langle r \rangle$}, of the nodes that fail at every iteration, divided by $\zeta$, with $ y = a/2$ marked in red, for two $\zeta$ values.
			For this figure $L = 4000$.}
		\label{fig: r_average2}
		
	\end{figure}
	
	In addition, when analyzing the cascading failures near the critical point in Fig. \ref{fig: r_average2}, we see that the behavior of $\langle r \rangle$ (the average distance from the center of the nodes that failed in every iteration) is consistent with our theory for our original model.
	We find that in the random branching process, $\langle r \rangle$ is constant for many iterations at distance $ \approx \frac{a}{2}\cdot\zeta$, but with the lower value of $a \approx 3.2$.
	And additionally, we see that in the spatial spreading process, the slope is fixed, meaning that the velocity of the cascading has linear dependent in $\zeta$, as in our original model.
	

\end{document}